# TRIGGERED ION ACOUSTIC WAVES IN THE SOLAR WIND

F.S. Mozer[1], I.Y. Vasko[1], and J. L. Verniero[1]

1 Space Sciences Laboratory, University of California, Berkeley, CA. 94720

I  ABSTRACT

For more than 12 hours beginning on January 18, 2021, continuous narrowband electrostatic emissions were observed on Parker Solar Probe near 20 solar radii. The observed <1000 Hz frequencies were well below the local ion plasma frequency. Surprisingly, the emissions consisted of electrostatic wave packets with shock-like envelopes, appearing repetitively at a ~1.5 Hz rate. This repetitiveness correlated and was in phase with low frequency electromagnetic fluctuations. The emissions were associated with simultaneously observed ion beams and conditions favorable for ion-acoustic wave excitation, i.e. $T_e/T_i$~5. Based on this information and on their velocity estimates of about 100 km/s, these electrostatic emissions are interpreted as ion-acoustic waves. Their observation demonstrates a new regime of instability and evolution of oblique ion-acoustic waves that have not been reported previously in theory or experiment.

II INTRODUCTION

Ion acoustic waves have been observed via spectral measurements on early satellites in the solar wind [Gurnett and Anderson, 1977; Gurnett and Frank, 1978; Kurth et al, 1979]. They are Doppler-shifted and measured at relatively high frequencies (from about the ion plasma frequency to the electron plasma frequency) in the spacecraft rest frame, due to their short wavelengths that scale with the local Debye length. The instabilities that produce these broadband electrostatic fluctuations might include the ion beam instability (also known as the ion-ion acoustic instability) [Lemons et al, 1979; Gary and Omidi, 1987], and electrostatic electron heat flux (electron-ion) instability [Forslund, 1970], but no consensus had been reached on the origin of these waves in the solar wind [Gurnett, 1991]. Ion-acoustic waves may be associated with ion beams that are produced by low-frequency turbulent magnetic field fluctuations [Valentini et al, 2009, 2011, 2014]. In this scenario, ion-acoustic waves provide thermalization of the ion beams, resulting in ion heating and termination of the turbulent cascade at short scales.

Recent data from the Parker Solar Probe has shown that Doppler shifted ion acoustic fluctuations are a dominant wave mode in the solar wind near the Sun [Mozer et al, 2020a,b]. Previous time-domain burst electric field measurements allowed analysis of electric field waveforms that showed the ion-acoustic fluctuations consisted of ion-acoustic wave packets (whose frequency changed from one packet to the next) [Mozer et al., 2020a] and nonlinear electrostatic structures interpreted as ion and electron phase space holes [Mozer et al., 2020b]. In spectral measurements, these waves and structures are observed as broadband wave activity above the local ion plasma frequency. In this Letter, we present intriguing observations of electrostatic wave emissions whose temporal behavior is drastically different from that observed in the earlier measurements. We argue that these measurements demonstrate a new regime of instability and evolution of ion-acoustic waves that has not been reported previously in observations or theory.

III DATA

The data presented in this paper comes from the magnetic and electric field measurements provided by the Parker Solar Probe FIELDS instrument [Bale et al. 2016], and electron and ion moments, and velocity distributions measured by the SWEAP instrument [Kasper et al., 2016]. The electric field experiment consists of four cylindrical booms in the spacecraft X-Y plane (the geometric half-length of each antenna pair is 3.5 m), which is perpendicular to the satellite-Sun line near perihelion (Bale et al. 2016). Data from the electric field experiment is transmitted at various rates and is presented in the spacecraft frame. The data rate of interest in the present work is ~2,200 S/s (Samples per second). Search coil magnetic field measurements were available at the same rate. In addition, electric field measurements in the burst mode were available for few second intervals at ~150,000 S/s resolution. The background magnetic field near perihelion was typically directed toward or away from the Sun, so the electric field components EX and EY measured by the Parker Solar Probe were typically components perpendicular to the background magnetic field.

Figure 1 presents spectra of electric and magnetic fields measured at 2200 S/s and obtained near the Parker Solar Probe's 20 solar radius perihelion at a time when the spacecraft was near to but not in the heliospheric current sheet. Panels 1B and 1C illustrate the spectra of the 500-1000 Hz electric field wave that existed for about 12 hours. This wave activity has no counterpart in the magnetic field spectra of the bottom three panels of the figure. The cross-spectrum of figure 1A shows that the phase difference between EX and EY was sometimes 180 degrees (the black or red parts of the curve) and sometimes 0 degrees (the green parts of the curve). These differences are due to the presence of more than a single wave and they show that all such waves were linearly polarized. The absence of a wave magnetic field and the linear polarization of the electric field require that this was electrostatic wave activity.  The background magnetic field of about 200 nT was directed predominantly along the Z-axis and towards the Sun, while measurements of the EZ electric field were not made. Therefore, we cannot directly determine the angle between the background magnetic field and the wave electric field, which is equivalent to the wave vector for electrostatic waves. However, the X-Y components of the wave vector electric field and the background magnetic field were measured to be not aligned, which is a strong indication that the electrostatic waves propagated obliquely to the local magnetic field.

To determine the nature of the electrostatic waves, their phase velocity is estimated from the data in Figure 2.  This figure presents the interferometry analysis of the electrostatic waves measured at 150,000 S/s, which frequency is sufficiently high to resolve the phase velocity of the waves. The top panel in Figure 2 presents EX and EY during a 20 ms interval. The quantities dV1, dV2, dV3, and dV4 in the two bottom panels of figure 2 are defined as follows. Let small letters, such as in v1, v2, v3, v4, or vsc, define the potential of the given antenna or the spacecraft body with respect to infinity.  The potentials that are actually measured are V1=v1-vsc, V2=v2-vsc, V3=v3-vsc, and V4=v4-vsc.  To remove the dependence of V1 through V4 on the potential of the spacecraft with respect to infinity, form

   dV1 ≡ V1-(V3+V4)/2 = (v1-vsc)-(v3-vsc)/2-(v4-vsc)/2 = v1 - (v3+v4)/2

Because (v3+v4)/2 is approximately the potential at the center of the spacecraft with respect to infinity, dV1 through dV4 represent the potentials of the antennas with respect to the center of the spacecraft. They do not depend on the potential of the spacecraft with respect to infinity, so they are the best measures of the desired quantities.

dV1 and dV2 are presented in the middle panel of figure 2 and dV3 and dV4 are given in the bottom panel.  In each case, one of the potentials leads the other by about 0.1 milliseconds (a fraction of a data point), which is the time for the wave to travel the ~two meter distance from the antenna to the spacecraft. This results in a wave speed estimate of tens of km/sec for the component of the wave speed in the X-Y plane.  This is the minor component of the wave speed (because the wave largely moves along B, which is in the Z direction), so the total wave speed is ~100 km/sec.  This is the order of an ion speed and not an electron speed. This is strong evidence that the wave is an ion-acoustic wave. The frequency variation of the wave, exhibited in figure 1A, 1B and 1C, is due to the Doppler shift of the wave being proportional to the Debye length actually measured during this time interval.

Figure 3 presents the ion velocity distributions and the bump-on-tail spectrum that was typical through the entire event. The temporal resolution of the ion distribution function is 1s. This distribution can be fit to core and beam Maxwellian populations with the following properties: core and beam densities of 1220 cm$^{-3}$ and 31 cm$^{-3}$; drift velocity between core and beam of -180 km/s (with the beam propagating anti-sunward faster than the core); perpendicular temperature of the core and beam of $T_c$~10 eV and $T_b$~17 eV; and temperature anisotropies $T_{perp}/T_{par}$ of core and beam of about 1.3 and 0.8. These conditions are favorable for the existence of ion-acoustic waves because the electron temperature of 50 eV was about five times greater than that of the dominant ion core population, i.e., $T_e/T_i$~5. This temperature ratio is not surprising because, in this time interval, the solar wind was very slow (200 km/s), which is known to correspond to electrons much hotter than ions (Salem et al., 2021 and references therein). The expected ion-acoustic speed is about $C_{ia}$ ~ $((T_e+3T_c)/m_i)^{1/2}$ ~ 120 km/s. Because the beam velocity is about 180 km/s, it is expected that the most unstable waves in this situation propagate obliquely at an angle ~45°. The facts that the waves are observed to have velocities of about 100 km/s and propagate obliquely to the magnetic field (in the XY plane) proves that the observed electrostatic emissions are ion-acoustic waves, produced most likely by the ion-ion beam instability. The observed parameters of the ion distribution function are around the marginal stability limit of the ion-acoustic instability [Gary and Omidi, 1987], provided the electron distribution function in the resonance regions (below a few eV, where the plasma instrument cannot measure electron distributions) does not cause strong damping.

It is desired to determine the properties of these waves in the plasma rest frame. The plasma density during the considered interval was about 2000 cm$^{-3}$, so the ion plasma frequency was about $f_{pi}$~9 kHz. The electron cyclotron frequency was about $f_{ce}$~6 kHz. In the spacecraft rest frame, the electrostatic waves had frequencies of f~500-1000 Hz, which is between 0.1 and 0.2 $f_{ce}$ and below 0.1 $f_{pi}$. The anti-sunward beam indicates that, in the plasma rest frame, the ion-acoustic waves should propagate anti-sunward. Therefore, the observed Doppler-shifted frequency, f, and the plasma rest frame frequency, $f_0$, are related as $2\Pi f = 2\Pi f_0 + kV_{sw}\cos\theta$, where θ is essentially the wave normal angle because the solar wind flow is thought to be parallel to the magnetic field within a few degrees. Because $f_0$<<$f_{pi}$, the ion-acoustic waves are almost dispersionless, so $2\Pi f_0$~$kC_s\cos\theta$ and the frequency of the waves in the plasma rest frame is f~ $f_0$ $c_s/(c_s+V_{sw})$. Because $V_{sw}$~200 km/s, the plasma rest frame frequencies of the observed waves were f~200-400 Hz, which is well below $f_{ce}$ and $f_{pi}$, and way above the ion cyclotron frequency of a few Hz. Because cosθ should be between 0.5 and 1 (the expected propagation angle is around 45º), the waves have wavelengths ~100-400 meters, which is 100-400 Debye lengths or a few thermal electron gyroradii.

Figure 4 presents snapshots of EX and EY over time intervals of 30 minutes to 50 milliseconds. The top panel gives a 30-minute example of the fact that the waves lasted several hours. Because EX was much larger than EY, the waves were predominately confined to the ecliptic plane. In addition, the wave amplitude had a spiky structure during all time intervals in the figure as well as over much of its 12-hour observation. In successive panels of figure 4, it is seen that the wave activity consisted of wave packets appearing repetitively at a frequency of about 1.5 Hz (fourth panel) and the two components have a phase difference of 180 degrees (bottom panel). Each wave packet has an unusual shock-like envelope with a duration of about 300 ms. Because ion-acoustic waves with $k\lambda_D \ll 1$ are weakly dispersive, the group velocity should be the sane order as the phase velocity and, hence, the envelopes have spatial scales of about 100 km.

To investigate the possible relationship of the wave repetition rate at 1.5 Hz and low frequency turbulence, the 0-5 Hz power spectra of the electric and magnetic fields are presented in figure 5 during the time interval of figure 1. That there were ~1.5 Hz electromagnetic waves throughout the interval suggests a possible relationship between the low frequency electromagnetic turbulence and the 500-1000 Hz ion acoustic wave repetition rate. This correlation is shown in figure 6, which presents a five second plot of EX that is high pass filtered at 0.5 Hz. The ~1.5 Hz waveform is seen as the low frequency sine wave in the figure while the 500-1000 Hz ion acoustic wave appears as the heavy black regions during the descending portion of each low frequency cycle. Thus, the high frequency ion acoustic wave repetition is phase correlated with the low frequency electromagnetic wave.

IV DISCUSSION

Observations are presented of electrostatic wave emissions observed continuously for about 12h in the solar wind aboard the Parker Solar Probe. These waves are interpreted as ion-acoustic waves because they are electrostatic and have phase velocities about equal to the local ion acoustic speed of 120 km/s. Another strong argument in favor of this interpretation is that, during this wave activity, an ion beam and electrons five times hotter than the core ions and about three times hotter than the beam ions were observed. Inspection of the Gary and Omidi, [1987] stability analysis shows that this system is at the marginal stability threshold of the ion-ion acoustic instability. The observed waves had quite low frequencies and long wavelengths; they had frequencies less than 0.05 $f_{pi}$ in the plasma rest frame and wavelengths of 100-400 Debye lengths or a few electron thermal gyroradii.

There are several unusual features of these ion-acoustic waves. The narrowband spectrum is the first intriguing feature. Ion beam instabilities produce quasi-monochromatic waves in the linear stage, but ion and electron trapping by the electrostatic waves makes them nonlinear and leads to a mixture of wave packets and isolated solitary structures like electron and ion holes [e.g., Børve et al., 2001; Muschietti and Roth, 2008]. In the nonlinear stage the spectrum finally becomes rather broadband as was earlier observed for ion-acoustic waves [Mozer et al., 2020a,b]. The shock-like envelopes and repetitiveness of the wave packets is the second intriguing features of the observed ion-acoustic waves. Neither of these features have been previously reported for the dynamics of the ion-acoustic waves and these results require a detailed theoretical analysis. The tentative scenario for this phenomenon is that the low-frequency waves make the marginally stable plasma locally unstable to emission of quasi-monochromatic ion-acoustic waves, which may explain the repetitiveness, while further competition between weak, dispersion, nonlinear steepening and particle trapping results in their evolution into shock-like envelopes.

It is noted that the electrostatic waves observed at ~200 Hz near the end of figure 1 also have ~1.5 Hz periodicities that are phase locked with the ~1.5 Hz electromagnetic waves, similar to the 500-1000 Hz waves. This shows that the triggering process can occur at more than one wave frequency at a given time.

The frequency range of 0.1-.0.2 fce and wave numbers the order of the thermal electron gyroradius are similar to those expected for oblique whistler waves driven by strahl electrons [Vasko et al., 2019; Verscharen et al., 2019]. Moreover, similarly the waves are oblique. Therefore, these waves can efficiently scatter electrons in pitch angle via anomalous and normal cyclotron resonances [Vasko et al., 2019; Verscharen et al., 2019]. The most efficiently scattered electrons will have energies of 50 eV - 500 eV, which is the strahl energy range. Thus, these ion acoustic waves might contribute to the electron heat flux regulation. Therefore, understanding their origin and effects is crucially important and is being addressed in follow-on studies.

V ACKNOWLEDGEMENTS

This work was supported by NASA contracts NNN06AA01C and NASA-G-80NSSC1. The work of I.V. was supported by National Science Foundation grant No. 2026680. I.V. also thanks the International Space Science

Institute, Bern, Switzerland for support. The authors acknowledge the extraordinary contributions of the Parker Solar Probe spacecraft engineering team at the Applied Physics Laboratory at Johns Hopkins University. Our sincere thanks to P. Harvey, K. Goetz, and M. Pulupa for managing the spacecraft commanding and data processing, which has become a heavy load thanks to the complexity of the instruments and the orbit. We thank the Fields Instruments P.I. S. Bale for scientific leadership of the mission. The data used in this paper are available at [http://fields.ssl.berkeley.edu/data](http://fields.ssl.berkeley.edu/data)

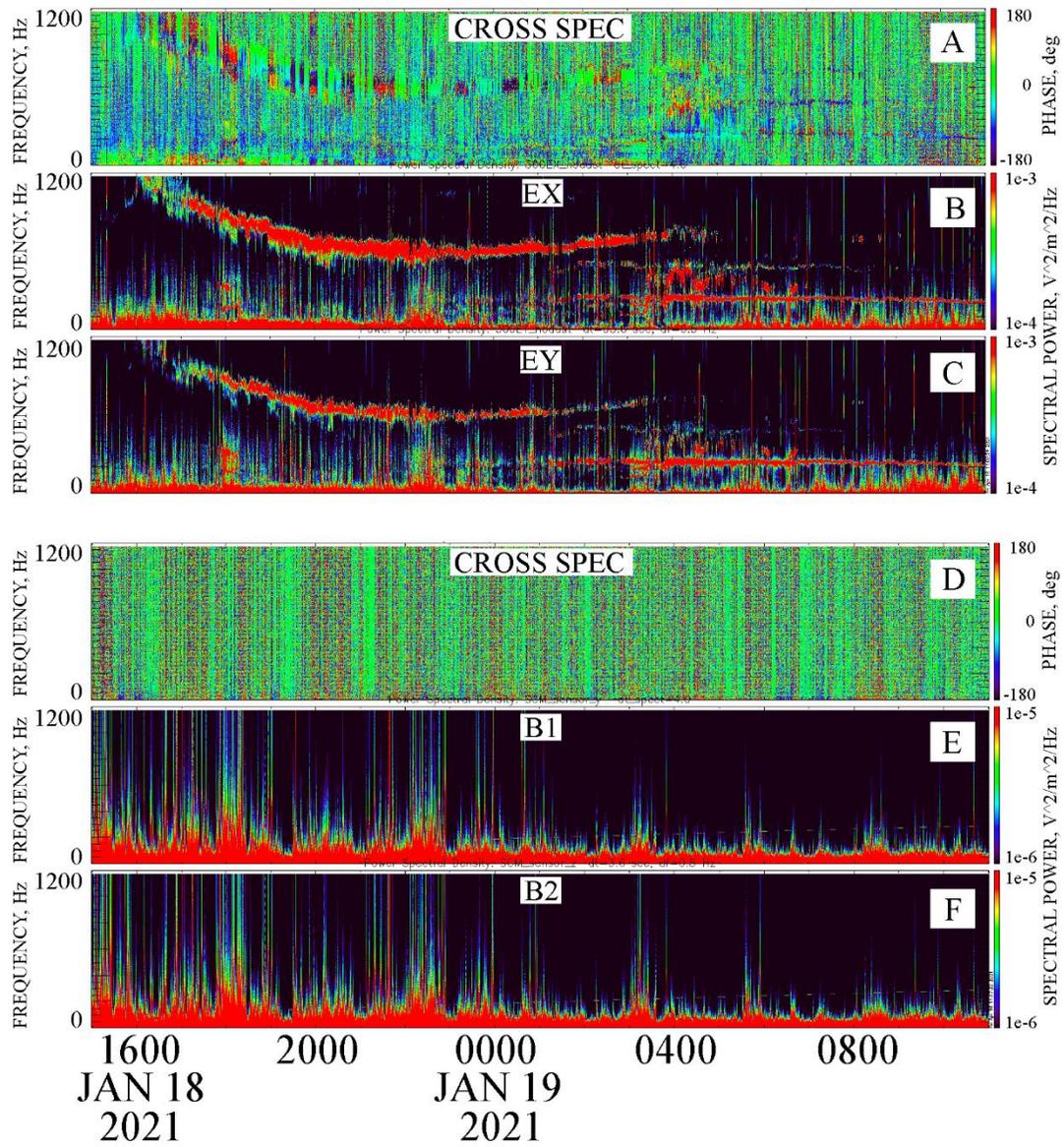

Figure 1. A twelve-hour period of continuous 600-1000 Hz waves seen in the electric field of panels 1A, 1B, and 1C but not in the magnetic field of panels 1D, 1E, and 1F. Also, note a continuous ~200 Hz electrostatic wave during the latter half of the interval.

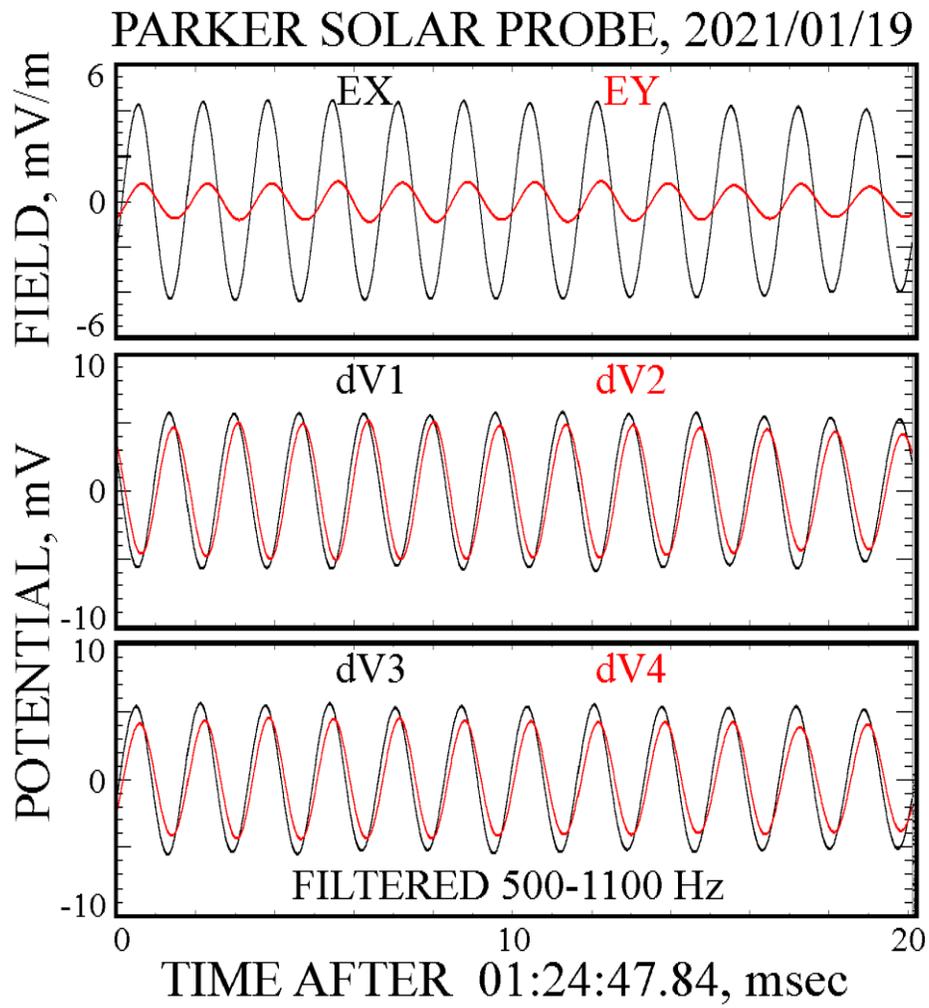

Figure 2. Electric field components (top panel) during a 20 millisecond interval along with potentials of the individual antennas (bottom panels), where the plotted quantities are defined in the main text. The ~0.1 millisecond differences between the individual antennas show that the wave speed was the order of an ion, not an electron, thermal speed.

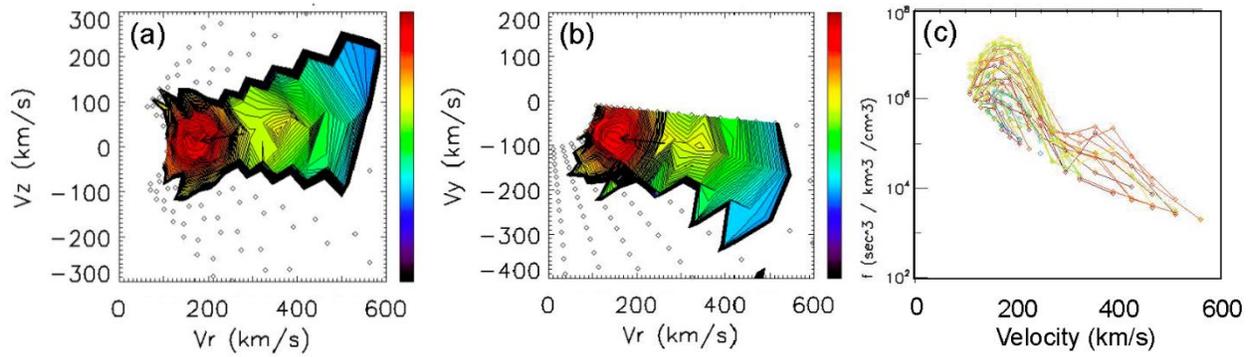

Figure 3. The ion velocity distribution functions, in panels (a) and (b), observed at 00:16:49 on January 19, 2021, and the spectra, in panel (c), which demonstrate the presence of ion beams and a bump-on-the-tail distribution

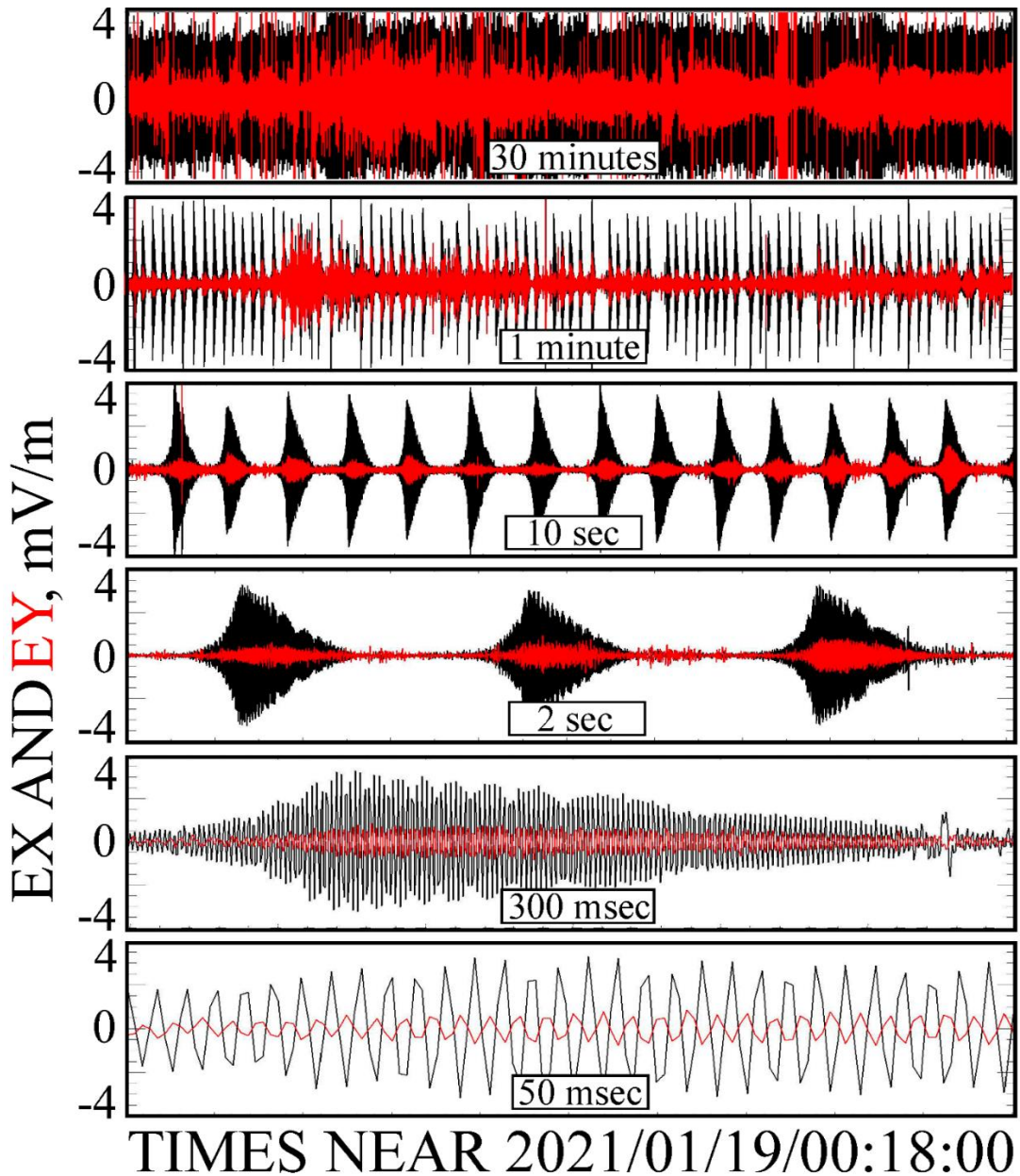

Figure 4. EX and EY waveforms, high pass filtered at 100 Hz, plotted over time intervals from 30 minutes to 50 milliseconds. As shown in the top panel, the pulsating wave existed for several hours. Because EX>>EY, the wave was largely confined to the ecliptic plane. The wave had a burst repetition rate of about 1.5 Hz and its two components were out of phase, which shows that the wave was linearly polarized.

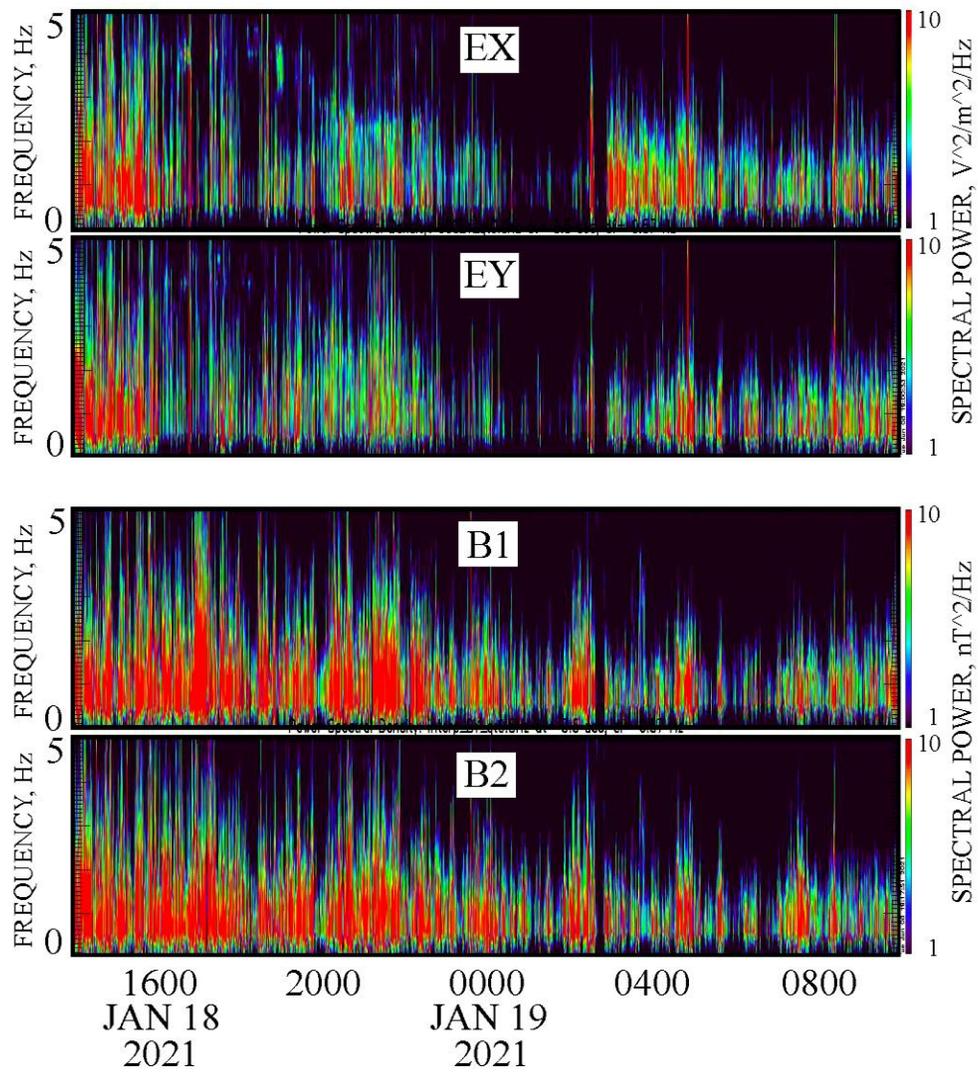

Figure 5. Spectra of E and B. They show the presence of a few Hz electromagnetic wave during essentially the entire time interval of figure 1.

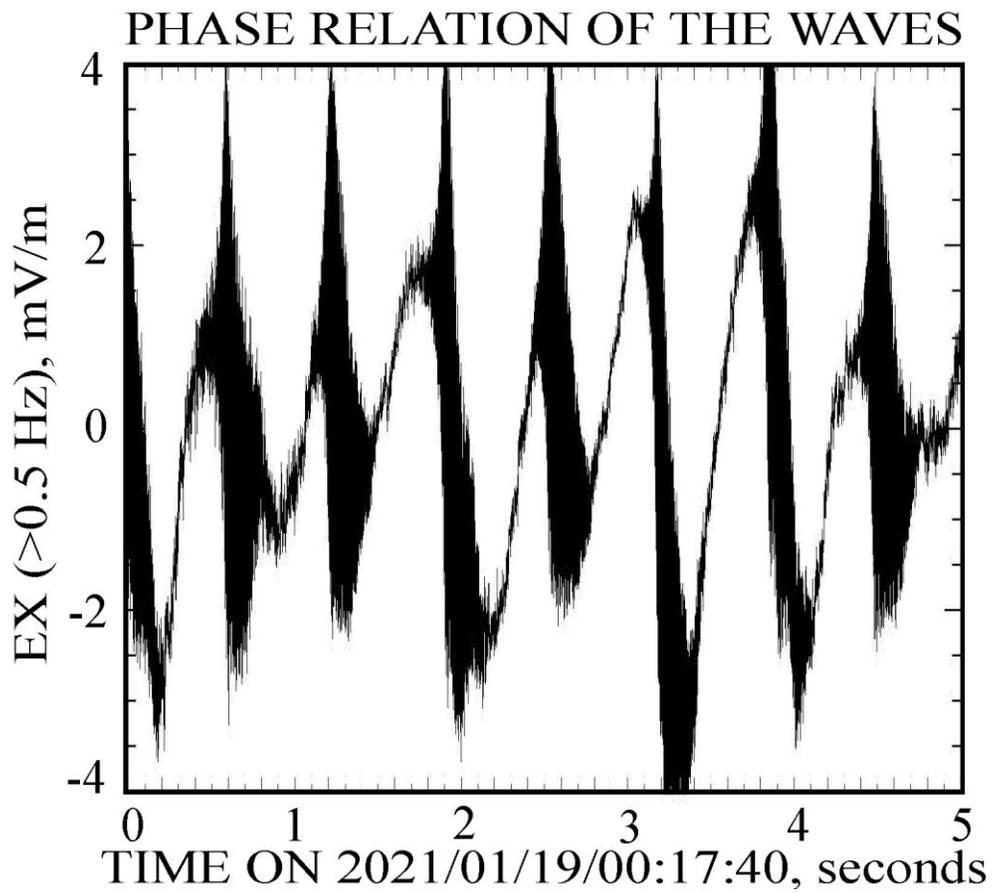

Figure 6. The EX electric field, high pass filtered above 0.5 Hz, which shows that the sine wave low frequency electromagnetic wave signature and the ~600 Hz electrostatic wave (the dark black regions) were phase correlated.